\documentclass[traditabstract]{aa}
\usepackage{natbib,twoopt,amssymb,lscape}
\usepackage[breaklinks=true]{hyperref} 
\bibpunct{(}{)}{;}{a}{}{,} 
\newcommandtwoopt{\citeads}[3][][]{\href{http://adsabs.harvard.edu/abs/#3}%
{\citealp[#1][#2]{#3}}} 
\newcommandtwoopt{\citepads}[3][][]{\href{http://adsabs.harvard.edu/abs/#3}%
{\citep[#1][#2]{#3}}} 
\newcommandtwoopt{\citetads}[3][][]{\href{http://adsabs.harvard.edu/abs/#3}%
{\citet[#1][#2]{#3}}} 
\newcommandtwoopt{\citeyearads}[3][][]%
{\href{http://adsabs.harvard.edu/abs/#3}{\citeyear[#1][#2]{#3}}}
\usepackage{longtable}
\usepackage{footnote}
\bibpunct{(}{)}{;}{a}{}{,} 
\usepackage{graphicx}
\usepackage{txfonts}

\makeatletter

\newcommand{\Rmnum}[1]{\expandafter\@slowromancap\romannumeral #1@}
\makeatother
\begin{document}
\begin{savenotes}
   \title{Symbiotic stars in X-rays \Rmnum{2}: faint sources detected with XMM- Newton and Chandra.}

   \author{N. E. Nu\~nez,\inst{1} 
   G. J. M. Luna, \inst{2}
            I. Pillitteri \inst{3}
            \and
            K. Mukai \inst{4}
          }
   \institute{Instituto de Ciencias Astron\'omicas de la Tierra y el Espacio (ICATE-UNSJ), Av. Espa\~na (sur) 1512, San Juan - Argentina \\
              \email{nnunez@icate-conicet.gob.ar}
         \and
             Instituto de Astronom\'ia y F\'isica del Espacio (IAFE), CC 67 - Suc. 28 (C1428ZAA)  CABA - Argentina.
         \and   
         Osservatorio Astronomico di Palermo (INAF), Piazza del Parlamento 1, 90\,134 Palermo - Italy.
          \and
          CRESST and X-ray Astrophysics Laboratory, NASA/GCFC, Greenbelt, MD 20\,771 - USA. Department of Physics, University of Maryland, Baltimore County, 1\,000 Hilltop Circle, Baltimore, MD 21\,250 - USA.\\
               }
   \date{}

  \abstract
   {We report the detection, with {\it Chandra} and XMM-{\it Newton}, of faint, soft X-ray emission from four symbiotics stars that were not known to be X-ray sources. These four object show a $\beta$-type X-ray spectrum, i.e. their spectra can be modeled with an absorbed optically thin thermal emission with temperatures of a few million degrees. Photometric series obtained with the Optical Monitor on board XMM-{\it Newton} from V2416 Sgr and {NSV 25735} support the proposed scenario where the X-ray emission is produced in a shock-heated region inside the symbiotic nebulae.} 

    \keywords{binaries: symbiotic, X-rays: individuals: Hen~2-87, NSV 25735, V2416~Sgr and Hen~2-104}
    \titlerunning{Symbiotic stars in X-rays 2}
    \maketitle

\section{Introduction} 
\label{sec:intro}

In a binary system where a white dwarf (WD) accretes from the wind of a red giant companion and frequently forms an accretion disk; where the WD can experiences frequent outburst of different intensity and where mass is ejected at speeds of a few hundreds km s$^{-1}$, it seems natural to expect that X-ray emission could be frequently detected, originated either in the surface of the white dwarf by quasi-stable nuclear burning, in the accretion disk, or in shocks due to ejections of material.
However, it was only recently that a significant fraction of these systems, known as WD symbiotics, was detected at X-ray wavelengths using modern instruments such as {\it Swift}, XMM-{\it Newton} or {\it Chandra}. In general, the emission is faint, with fluxes as low as a few 10$^{-14}$ ergs cm$^{-2}$ s$^{-1}$. WD symbiotics have X-ray spectra that can be classified into four types: supersoft X-ray sources whose X-ray spectrum peaks at energies lower than 0.4 keV, named $\alpha$-type; $\beta$-type comprises those sources with soft X-ray emission and whose spectrum extends up to energies of less than 2.4 keV;  $\delta$-type are highly absorbed, hard X-ray sources with detectable thermal emission above 2.4 keV and $\beta/\delta$-type includes those WD symbiotics with two X-ray thermal components, soft and hard \citepads[this classification scheme is detailed in][hereafter Paper I]{2013A&A...559A...6L}.

In this paper we present the detection at X-ray energies of four WD symbiotics: \object{Hen~2-87}, \object{NSV 25735}, \object{V2416~Sgr} and \object{Hen~2-104} observed with XMM-{\it Newton} and {\it Chandra} that were not known to be X-ray sources before. In Section \ref{sec:analisis} we detailed the data reduction while in Section \ref{sec:results} we present the results, while discussion and concluding remarks are presented in Section \ref{sec:concl}. 

\section{Observations and data reductions}
\label{sec:analisis}

We searched for imaging X-ray observations sensitive in the 0.3-10 keV band ({\it ASCA}, XMM-{\it Newton}, {\it Chandra}, {\it Suzaku} and {\it Swift}) using the HEASARC\footnote{http://heasarc.gsfc.nasa.gov/docs/archive.html} database of all the symbiotic stars listed in \citepads{2000A&AS..146..407B}. Excluding the symbiotic stars already known to be X-ray sources (see Paper I which contains a complete list of known X-ray emitting symbiotic stars), we detected X-ray emission from four symbiotic stars: \object{Hen~2-87}, \object{NSV 25735}, \object{V2416~Sgr} and \object{Hen~2-104}.

\begin{table*}
\caption{Observations Log}
\label{table:info1}
\begin{tabular}{lccccc}
\hline\hline
Object & Instrument & ObsID & Date & Exposure time [ks] & Offset[\arcsec] \footnotemark[1] \\
\hline\
Hen~2-87         & XMM-{\it Newton} - EPIC & 0109480101 & 2\,002/06/03 & 53 & 10.402\\
                 &                               & 0109480201 & 2\,002/08/26 & 55 & 10.402\\
                 &                               & 0109480401 & 2\,003/01/21 & 48 & 10.403\\          
NSV 25735  & XMM-{\it Newton} - EPIC & 0401670201 & 2\,007/04/30 & 10 & 0.007 \\
V2416~Sgr        & XMM-{\it Newton} - EPIC & 0099760201 & 2\,000/03/22 & 61 & 0.005 \\
                 & XMM-{\it Newton} - OM         & 0099760201 & 2\,000/03/22 & 47 & 0.005  \\
Hen~2-104        & {\it Chandra} - ACIS    & 2577       & 2\,002/04/08 & 20 & 0.022 \\
\hline
\end{tabular}
\footnotetext{Offset from instrument aim point, obtained from HEASARC}

\end{table*} 

\object{Hen~2-87}, \object{NSV 25735} and \object{V2416~Sgr} were observed with XMM-{\it Newton}. While \object{NSV 25735} and \object{V2416~Sgr} were the main target of the observations, \object{Hen~2-87} was serendipitously detected at the edge of the field of view of the EPIC camera, about 10\arcsec away from the aim point. The XMM-{\it Newton} data were reprocessed using the SAS (v13.0.1) tool \texttt{emproc} and \texttt{epproc} to apply the latest calibration. All observations were obtained in full mode and with the thick filter. After filtering the event list for flaring particle background, Hen~2-87 and V2416~Sgr were evidently detected and we extracted source spectra and light curves from circular regions centered on the SIMBAD coordinates of each objects with radius of 600 or 640 pixels ($\sim$32\arcsec) (we used an smaller region in the case of Hen~2-87 because it lies close to a detector edge). Ancillary and response matrices were created with the SAS tools \texttt{arfgen} and \
texttt{rmfgen}. The background spectra and light curves of NSV 25735 and V2416~Sgr were extracted from annulii regions centered on the source with inner and outer radii of 650 and 1\,000 pixels, respectively. Because Hen~2-87 is located near the detector edge, we extracted the background from a circle off-centered from the source region with 90\arcsec radius. Given the faintness of the sources, we combined the $pn$, $mos1$ and $mos2$ spectra (using the \texttt{epicspeccombine} script) increasing the signal-to-noise of the spectra to be modeled. 

To search for X-ray emission from \object{NSV 25735}, we used an algorithm based on wavelet convolution for the EPIC XMM-{\it Newton} images. This code is derived from the  {\it ROSAT} and {\it Chandra} versions (see \citeads {1997ApJ...483..350D} and \citeads {1997ApJ...483..370D} for details). The XMM-{\it Newton} version allows to perform source detection on the data of the three EPIC camera simultaneously, improving thus the sensitivity of the result. The detected sources are those with a significance threshold higher than 4.5 $\sigma$ of the local background. This threshold is chosen on the basis of simulations of background only images, in order to retain at most one spurious source in the image. \object{NSV 25735} is detected at a significance level of 15.22 $\sigma$, it has about 400$\pm 40$ counts in the sum of $pn$, $mos1$ and $mos2$ images and a rate of $\sim 12$ counts s$^{-1}$.

NSV 25735 and V2416~Sgr were also observed with the optical monitor (OM) onboard XMM-{\it Newton}. While NSV 25735 was observed in fast imaging mode with the UVM2 filter, V2416~Sgr was observed in image mode using the U, B and V filters (UVW2 and UVM2 filters were also used but the data were corrupted). We analized pipeline-reduced data and look for variability in the photometric time series. NSV 25735 has been observed with {\it Swift}/XRT/UVOT, however, the source was not detected in X-rays and the UVOT images were saturated (see Paper I for details). 

\object{Hen~2-104} was observed with {\it Chandra} using the ACIS-S camera. We reprocessed the data using the \texttt{chandra$\_$repro} script in CIAO\footnote{http://cxc.cfa.harvard.edu/ciao/index.html} and obtained a new event file with the calibration version 4.5.2 applied. Afterwards, we extracted spectra (using the \texttt{specextract} script) and light curves (using the \texttt{dmextract} script) for the source and background. Source products were extracted from a circular region (centered on the SIMBAD\footnote{http://simbad.u-strasbg.fr/simbad/sim-fid} coordinates) with 100 pixels ($\sim$50\arcsec) radius while background spectra and light curves were extracted from 3 circular regions of 160 pixels ($\sim$ 80\arcsec) each. The response and ancillary matrices were created using the \texttt{mkrmf} and \texttt{mkarf} scripts. All the observations are detailed in Table \ref{table:info1}. 

We fit the unbinned spectra using XSPEC\footnote{http://heasarc.gsfc.nasa.gov/docs/xanadu/xspec/} and because it is not appropriate to use the $\chi^{2}$ statistic to fit the spectra of sources with a low numbers of counts, such as the ones studied here, we use the C-statistic \citepads{1979ApJ...228..939C}. All errors in the fit parameters were estimated at their 90\% confidence. To evaluate how many components are needed in the spectral model, we use the likelihood ratio test as described by \citetads{2002ApJ...571..545P} and implemented in XSPEC through the {\texttt lrt} script. This test compares the statistics (Cstat) of the complex and simple model in $N$ sets of simulated data and helps us to evaluate how statistically significant is to add one or more components into a single-component model.

\section{Results}
\label{sec:results}
 
By correlating the existing catalogue of \citetads{2000A&AS..146..407B} with the HEASARC database, we found X-ray emission from four symbiotic stars observed with XMM-{\it Newton} and {\it Chandra}. Their low X-ray fluxes (see Table 2) explain why these sources were not detected before. Given the limited statistical quality of these X-ray data, we limit ourselves to spectral models that have been successfully used for other symbiotic stars. All of them show a soft optically thin thermal X-ray spectrum with emission that peaks around 1 keV and do not extend to very high energies. This is sufficient to fit their spectra and classify them as $\beta$-type in the scheme detailed in Paper I (in the next subsections we detail the fit procedure for each object). We also tried to fit the X-ray spectrum with non-thermal models which were discarded because they yielded unrealistic values for the fit parameters.

For all sources, we show the classification based on infrared (IR) colors proposed by \citetads{1975MNRAS.171..171W} and refined by \citetads{1984PASAu...5..369A}, where $S$-type systems emit IR radiation typical of red giant atmospheres (M-type), they are relatively dust-free and have binary periods shorter than 20 years; $D$-type systems show IR emission indicative of dust, that is, thermal radiation with average temperatures of 1\,000 K, generally contain Mira variables as companions and have binary periods longer than 10 years like R Aqr who show a period of 46 years \citepads{2009A&A...495..931G}. $D'$-type systems have very red colors in the far IR and a cool companion of spectral types F or G. The IR type of all objects was extracted from Table 1 in \citetads{2000A&AS..146..407B}.

\begin{figure*}
\centering
\label{fig:spec}         
     \includegraphics[]{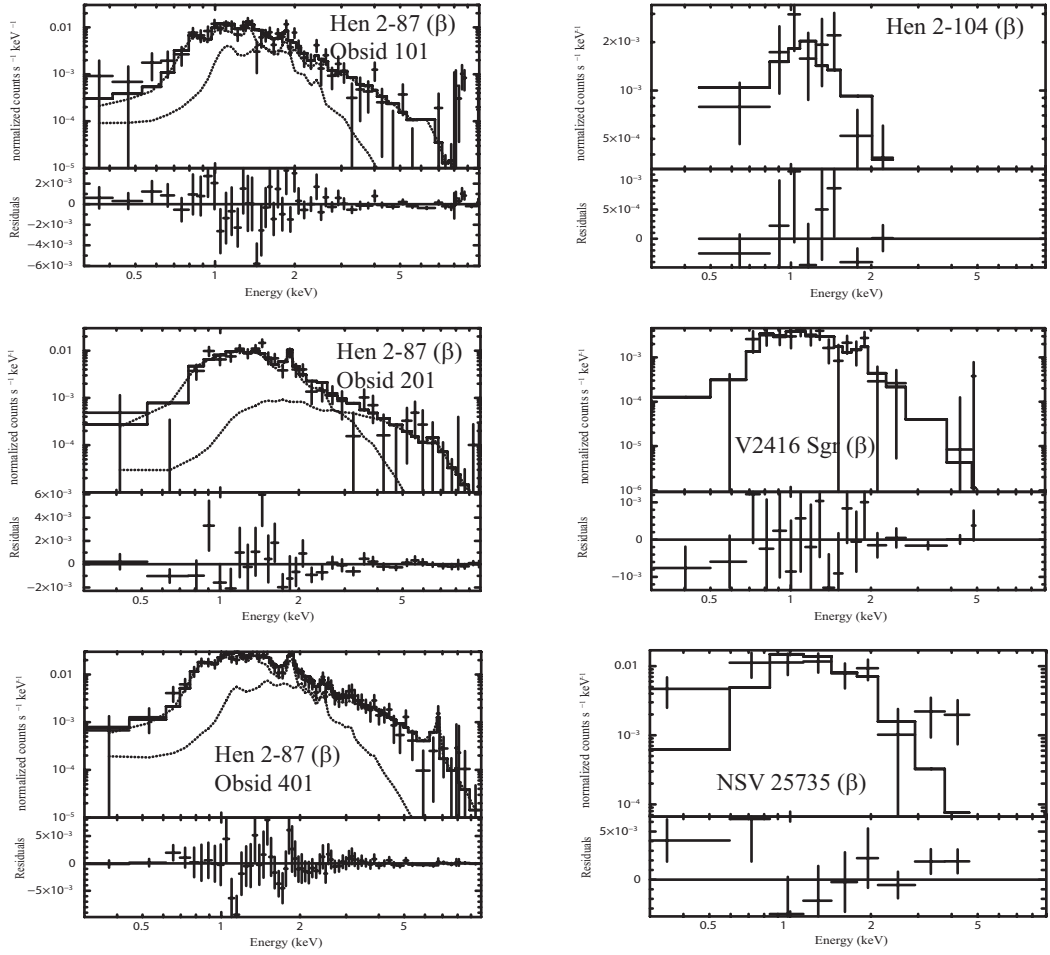}
     \caption{{\it Chandra} and XMM-{\it Newton} X-ray spectra of the WD symbiotic with newly discovered X-ray emission together with their X-ray spectral types: Hen~2-87, Hen~2-104, V2416~Sgr and NSV~25735. We show the 3 spectra obtained with XMM-{\it Newton} from Hen 2-87. The full line shows the best-fit model described in Section \ref{sec:results}, while the dotted line shows the contribution of the individual spectral components in the case of multi-component models. The X-ray spectral classification for each source is included between parentheses in each panel.}
 \end{figure*}

\subsection{Hen~2-87}
\label{sec:hen287}

This object is classified as $S$-type. Little is known about this source and there are not recorded outburst in the literature. \object{Hen~2-87} was detected in three XMM-{\it Newton} observations in the field of \object{WR 47} (see Table \ref{table:info1}). Given that the source was detected at a higher count rate during ObsID 0109480401 (hereafter 401) taken in 2003 (0.037$\pm$0.001  s$^{-1}$, compared to 0.011$\pm$0.002 counts s$^{-1}$ during ObsID 0109480201 (hereafter 201) and 0.013$\pm$0.002 counts s$^{-1}$ for ObsID 0109480101 (hereafter 101)), we fit this spectrum and then used the accepted model to fit the other two datasets. We first fit the spectrum with an absorbed optically thin thermal plasma (\texttt{wabs$\times$apec}) but there were important residuals in the $E \gtrsim$ 2 keV region. Then, we added a second optically thin thermal component, also affected by the same absorber than the previous component (\texttt{wabs$\times$(apec$_1$+apec$_2$)}, and the fit showed a much more uniform 
distribution of the residuals. As explained in Section \ref{sec:analisis}, we used the \texttt{lrt} to assess the statistical significance of adding this second component and after 1\,000 simulations, 60\% of them showed an improvement in the statistic. The addition of the second component has a low significance, however given that this component describe the high energy part of the spectrum, we find plausible to keep it in our model (see Fig. \ref{fig:spec}). We retain this complex model throughout the analysis, which could be better tested in more sensitive observations. During the 3 observations, the plasma temperatures and the absorbing column have commensurate values. The flux on the other hand, almost doubled from ObsID 101 to 401 (see Table 2). 

\subsection{NSV 25735}
\label{sec:NSV25735}

The symbiotic NSV 25735 is classified as $D'$-type. The system contains a WD with moderate temperature (T$\sim$50\,000 K,  Schmid \& Nussbaumer \citeyearads{1993A&A...268..159S}). The cool companion star has three different spectral type determination in the literature: G5 \citepads{1999A&AS..137..473M}, G4 III/IV \citepads{2001ApJ...556L..55S}, and G7 III \citepads{2001A&A...369L...1M} suggesting a T$_{eff}$ in a range between 5\,025-5\,300 K. The cool component is rotating at a spectacular speed of 105 km s$^{-1}$, unseen in field G-type giants \citepads{2001A&A...369L...1M}. As only 7 $D'$-type symbiotic are known, the fact that the 3 sources (HD~330036, AS~201 and NSV~25735) studied so far with high resolution optical spectroscopy show rapid rotation and $s$-process elemental overabundances, might indicate that these two traits are signatures of these symbiotic systems \citepads{2005A&A...441.1135J}. Moreover, NSV~25735 is the first $D'$-type symbiotic to show jets. \citetads{2001A&A...369L...1M} 
analyzed the H$\alpha$ profile and found a central component (that remained constant for months) and two weaker and symmetrically placed components at both sides of the profile. These two components show large day-to-day variability (velocity and width), which are spectral signatures of jet-like discrete ejection events. Also, the authors derived a high orbital inclination of 60$^{\circ}$, therefore the de-projected velocity of the jet components must be much larger than the observed velocity shifts (150 km s$^{−1}$) and well in excess of the escape velocity from the companion (1\,000 km s$^{−1}$). The jet components are visible in the profiles of He I lines as well, with a velocity of 300 km s$^{-1}$. There are not recorded outbursts but the bipolar mass outflow is highly variable \citepads{2001A&A...369L...1M}. The very fast rotation as been explained by \citetads{1996MNRAS.279..180J}, who proposed a mechanism in which the accretion of a slow massive wind from the AGB progenitor of the current WD can 
transfer sufficient angular momentum, spinning up the companion, in analogy with millisecond radio pulsars \citepads{1984A&A...139L..16V}.  This also explains the chemical enrichment in $s$-process elements in ${D'}$-type symbiotic systems that were present in the AGB wind \citepads{2005A&A...429..993P}.

The X-ray spectrum obtained with XMM-{\it Newton}, that extends up to energies of $\sim$ 5 keV, can be modeled using an absorbed ($N_{H}$=1.2$_{-0.4}^{+0.4}\times$10$^{22}$ cm$^{-2}$) optically thin thermal plasma with a temperature $kT$ = 0.7$_{-0.3}^{+0.3}$ keV. XMM-{\it Newton} detected 351 source counts, therefore the model parameters cannot be constrained by the fit (see Table 2). The unabsorbed flux in the 0.3-10.0 keV energy band is F$_{X}$=9$_{-3}^{+3} \times$10$^{-13}$ ergs cm$^{-2}$ s$^{-1}$ and the corresponding luminosity at a distance of 575 pc \citepads{2001A&A...369L...1M} is L$_{X}$=3$^{+2}_{-1} \times$10$^{31}$ ergs s$^{-1}$.

\subsection{V2416~Sgr}

This object was classified as a $S$-type symbiotic. XMM-{\it Newton} detected 220 counts from this source. The X-ray spectrum can be modeled with an absorbed ($N_{H}$=1.1$_{-0.2}^{+0.3}\times$10$^{22}$ cm$^{-2}$) optically thin thermal plasma (with kT = 0.5$_{-0.2}^{+0.2}$ keV). It is interesting to note that the value of $N_{H}$ that could be infered \citepads[using the relation between $N_{H}$ and E(B-V) proposed by][]{1989A&AS...79..359G} from the published reddening values of E(B-V)=1.7 is 4.4$\times$10$^{22}$ cm$^{-2}$ \citepads{2005A&A...435.1087L} or using E(B-V)=2.5 \citepads{1997A&A...327..191M} we have $N_H$=6.6$\times$10$^{22}$ cm$^{-2}$. Both values are much higher than the value found from our fit of the X-ray spectrum. On the other hand, \citetads{1992A&A...255..171W}, calculated a lower value of E(B-V)$\sim$0.13, but at the same time they mentioned that the colors of this star suggest it experiences several magnitude more extinction than the value derived. This suggest that the X-ray emitting 
region could be located in the outer parts of the symbiotic nebulae, where column density is smaller and so is the absorption. The unabsorbed flux in the 0.3-10.0 keV energy band is F$_{X}$=0.9$_{-0.2}^{+0.2} \times$10$^{-13}$ ergs cm$^{-2}$ s$^{-1}$ and the corresponding luminosity at a distance of 1.0 kpc (the actual distance is unknown) is L$_{X}$=1.0$^{+0.2}_{-0.2} \times$10$^{31}$ ergs s$^{-1}$.

The OM light curves from V2416~Sgr and {NSV~25735} show a small fractional variability $\lesssim$ 3\% (which we calculate as the observed variance over the average count rate), which is expected for their $\beta$ X-ray spectral type. In the diagram presented in Paper I (Fig. 7), where the hardness of the X-ray spectrum is plotted against the fractional UV variability, it can be clearly seen a separation between those systems that are accretion-powered and those that are powered by quasi-stable shell burning. \object{NSV 25735} and V2416~Sgr are located in the lower left corner, in the region of systems where the WDs are powered by quasi-stable shell burning.

\subsection{Hen~2-104}
\label{sec:hen2104}

\object{Hen~2-104} is classified as a $D$-type symbiotic. Its extended nebula has an hourglass shape, extending 75\arcsec \citepads{1989ApJ...344L..29S}. More detailed studies showed that the nebula around Hen~2-104 consists in nested hourglass-shaped structures (with the inner structure extending $\sim$ 12\arcsec while the outer structure extends for $\sim$ 80\arcsec) and a collimated polar jet \citepads[see Fig.1 in][]{2001ApJ...553..211C}. The outer pair of lobes and jets are produced by a high-velocity outflow from the WD companion of the Mira, while the inner, slowly expanding lobes are produced by the Mira wind itself \citepads{2001ApJ...553..211C}. The densities range from $n_{e}$=500 cm$^{-3}$ to 1\,000 cm$^{-3}$ in the inner lobes, and from 300 cm$^{-3}$ to 500 cm$^{-3}$ in the outer lobes \citepads{2008A&A...485..117S}.

The {\it Chandra} observation was already studied by \citetads{2006AAS...209.9206M}, although due to poor astrometry, they mistakenly attributed the X-ray emission to a region heated by shocks and displaced 2{\arcsec} SE from the central symbiotic system (R. Montez private communication). Our new data analysis shows that the X-ray emission is concentrated in the central region of the nebula. The faint X-ray emission can be modeled with an absorbed ($N_{H} \lesssim$ 0.3$\times$10$^{22}$ cm$^{-2}$) optically thin thermal plasma (with kT $\gtrsim$ 1.5 keV), however, due to the low number of counts detected (42) we are not able to constrain further the value of the model parameters, upper or lower limits are listed in Table 2. The unabsorbed flux in the 0.3-10.0 keV energy band is F$_{X}$=0.3$_{-0.1}^{+0.1} \times$10$^{-13}$ ergs cm$^{-2}$ s$^{-1}$ and the corresponding luminosity at a distance of 3.3 kpc \citepads{2008A&A...485..117S} is L$_{X}$=4.2$^{+1.0}_{-1.5} \times$10$^{31}$ ergs s$^{-1}$.

\begin{table*}
\label{table:xrays}
\caption{X-ray spectral fitting results. X-ray flux (F$_X$) and luminosity (L$_{X}$), in units of 10$^{-13}$ ergs s$^{-1}$ cm$^{-2}$ and 10$^{31}$ ergs s$^{-1}$ respectively, are calculated in the 0.3-10.0 keV energy band.}
\begin{tabular}{lcllcl}
\hline\hline
Object        & Model    & kT [keV]      & $N_H$ [10$^{22}$ cm$^{-2}$] & F$_X$ & L$_{X}$  \\
\hline
Hen 2-87(101) & \texttt{wabs$\times$(apec$_{1}$+apec$_{2}$)} & 1: 0.4$_{-0.2}^{+0.4}$ & 1.3$_{-0.2}^{+0.2}$  & 7.5$_{-0.9}^{+0.9}$ & 9.0 $_{-0.7}^{+0.7}$\\
              &                                              & 2: 1.9$_{-0.4}^{0.9}$ &  &  &  \\
Hen 2-87(201) & \texttt{wabs$\times$(apec$_{1}$+apec$_{2}$)} & 1: 0.7$_{-0.1}^{+0.2}$ & 1.2$_{-0.2}^{+0.2}$ & 8.1$_{-1.1}^{+1.1}$ &  9.7$_{-1.3}^{+1.3}$ \\
              &                                              & 2: $\gtrsim$2.0 &  &  & \\
Hen 2-87(401) & \texttt{wabs$\times$(apec$_{1}$+apec$_{2}$)} & 1: 0.6$_{-0.1}^{+0.1}$ & 1.4$_{-0.1}^{+0.1}$ & 14$_{-1}^{+1}$ & 20.0$_{-4.4}^{+2.2}$  \\
&  & 2: 3.0$_{-1.0}^{+1.0}$ &  &  &   \\
\object{NSV 25735} & \texttt{wabs$\times$apec} & 0.7$_{-0.3}^{+0.3}$ & 1.2$_{-0.4}^{+0.4}$ & 9$_{-3}^{+3}$ & 3$_{-1}^{+2}$  \\
&  &  &  &  &   \\
V2416~Sgr & \texttt{wabs$\times$apec} & 0.5$_{-0.2}^{+0.2}$ & 1.1$_{-0.2}^{+0.3}$ & 0.9$_{-0.2}^{+0.2}$ & 1.1$_{-0.3}^{+0.2}$  \\
&  &  &  &  &   \\
Hen 2-104     & \texttt{wabs$\times$apec} & $\gtrsim$1.5 & $\lesssim$0.3 & 0.3$_{-0.1}^{+0.1}$ & 4.2$^{+1.0}_{-1.5}$  \\
&  &  &  &  &   \\
\hline
\end{tabular}
\end{table*}
  

\section{Discussion and conclusions}
\label{sec:concl}

We found X-ray emission from four WD symbiotic that were not known to be X-ray sources. The X-ray emission from all of them can be classified as $\beta$-type, following the scheme originally proposed by \citetads{1997A&A...319..201M} and recently updated in Paper I. Our results confirm that $\beta$-type X-ray emission is the most frequent among WD symbiotic, increasing the number of known $\beta$-type sources to 16 out of 48 symbiotic with X-ray emission. The X-ray emission from the new sources is faint, with luminosities on average of 10$^{31}$ ergs s$^{-1}$ (albeit the uncertainties in their distance), which explains why pointing, dedicated observations were necessary for their detection.  As $\beta$-type systems, their X-ray emission seems to be associated with shocks in the nebulae that could be originated in a colliding wind region or inside extended jets (known to be present in \object{NSV 25735} and \object{Hen~2-104}).

Although most models of colliding-wind scenarios require that the WD in the system underwent a recent outburst to be able to trigger a fast, tenuous wind, the $\beta$-type systems presented here as well as those presented in Paper I do not have any recent outburst recorded. While $\alpha$ or $\delta$ type X-ray spectra are directly linked to the source that powers the WD, i.e. shell-burning or accretion, $\beta$ type X-ray emission is not. The absence of strong variability, together with a $\beta$-type X-ray spectrum strongly suggest that the observed X-rays are not originated in the accretion disk around the WD, while at the same time, those soft X-rays are too energetic to be originated in a shell-burning white dwarf. Our physical picture is that, in these systems, nuclear burning could lead to a strong wind from the WD (or the accretion disk near the WD), which collides with the wind from the cool star, leading to observed soft X-ray emission. Super-soft X-ray emission from the WD might be present as well 
but absorbed by the dense symbiotic nebula.

Using the temperature obtained from spectral models of these sources (0.4 keV to $\sim$2.0 keV) and assuming strong shock conditions, we can derive a shock speed, v$_{shock}$. We obtain v$_{shock}$ in the range of $\sim$1\,000 to $\sim$2\,500 km s$^{-1}$. Similar speeds are observed in the $\beta$-type sources presented in Paper I and \citetads{1997A&A...319..201M} and are roughly consistent with the speeds of X-ray emitting outflows from WD symbiotic \citepads{2007ApJ...660..651N,2004ApJ...613L..61G}. High outflow speeds of a few thousands km s$^{-1}$ have been observed in the optical line profiles of MWC~560 \citepads{2001A&A...377..206S}, although they are not consistent with the temperatures observed from the $\beta$ component in its X-ray spectrum \citepads{2009A&A...498..209S}. The disparity between X-ray infered speed and speeds from optical line widths has been also observed in the diffuse X-ray emission from planetary nebulae \citepads{2003ApJ...583..368S}. In the dense symbiotic nebula, however, 
some authors have modeled the observed UV line profiles of O~VI and He~II as due to electron scattering \citepads{2012BaltA..21..196S,2012MNRAS.427..979S}. \citetads{2000ApJ...541L..25L} proposed that Ramman scattering might be responsible for the broad H$\alpha$ line wings observed in symbiotics.

The number of known symbiotic with X-ray emission is rapidly increasing with the new instruments available and it is expected to increase even more with much sensitive instruments and deeper surveys (e.g. {\it NuSTAR}\footnote{http://www.nustar.caltech.edu/} and {\it eROSITA}\footnote{http://www.mpe.mpg.de/eROSITA}). We now know 48 systems with X-ray emission originating in four different scenarios. With this number of sources it seems timely to study their statistical properties. Is there any trend or relationship between IR and X-ray spectral types?. In the symbiotic stars catalog of \citetads{2000A&AS..146..407B}, 63 \% of the objects (139) are classified as $S$-type, 15\% (34) as $D$-type, and 3\% (7) as $D'$-type. Among the 48 symbiotic with X-ray emission, 33\% of them have a $\beta$-type X-ray spectrum. Roughly half of the $\beta$-type symbiotics have infrared $S$-type spectrum, while 25\% have $D$-type, 6.3\% have $D'$-type and 6.3\% do not have 
classification. We did not find any clear tendency or correlation between IR and X-ray types, i.e. the fractions of IR types are roughly similar among X-rays emitting symbiotics and those with no X-ray emission detected.

\begin{acknowledgements}

We thank Jennifer L. Sokoloski for useful discussions about X-ray emission from symbiotic stars. N. E. N. acknowledge Consejo Nacional de Investigaciones Cient\'ificas y T\'ecnicas, Argentina (CONICET) by the Postdoctoral Fellowship. G. J. M. Luna acknowledge funding from grants PICT/2011/269 from Agencia and PIP D-4598/2012 from Consensus National de Investigation Cient\'ificas y T\'ecnicas, Argentina. This research has made use of data obtained from the High Energy Astrophysics Science Archive Research Center (HEASARC), provided by NASA's Goddard Space Flight Center; and the VizieR catalogue access tool, CDS, Strasbourg, France. The original description of the VizieR service was published in \citetads{2000A&AS..143...23O}.
\end{acknowledgements}

\bibliographystyle{aa}    
\bibliography{listaref}
\end{savenotes}
\end{document}